\documentstyle[preprint,aps, epsfig]{revtex}

\begin{document}


\preprint{\hfill$\vcenter{\hbox{\bf IFT-P.076/99} }$ }

\title{ Single Production of Leptoquarks at the Tevatron }

\author{O.\ J.\ P.\ \'Eboli\thanks{Email: eboli@ift.unesp.br} and 
        T.\ L.\ Lungov\thanks{Email: thais@ift.unesp.br} }

\address{\em Instituto de F\'{\i}sica Te\'orica -- UNESP \\ 
             R. Pamplona 145, 01405-900 S\~ao Paulo, Brazil}

\maketitle

\vspace{.2in}


\hfuzz=25pt

\begin{abstract}
\vskip -18pt

 We study the single production of first generation leptoquarks in association
 with a $e^\pm$ at the Fermilab Tevatron. We focus our attention on final
 states exhibiting a $e^+e^-$ pair and jets, and perform a detailed analyses
 of signal and backgrounds.  The single leptoquark production cross section
 depends on the leptoquark Yukawa coupling to lepton-quark pairs and we show
 that the study of this mode can extend considerably the leptoquark search for
 a large range of these couplings. In fact, for Yukawa couplings of the
 electromagnetic strength, the combined results of the Tevatron experiments
 can exclude the existence of leptoquarks with masses up to 260--285
 (370--425) GeV at the RUN I (RUN II), depending on their type.

\end{abstract}


\newpage

\section{Introduction}

Leptoquarks are an undeniable signal of physics beyond the standard model
(SM), and consequently, there have been several direct searches for them in
accelerators.  In fact, many theories that treat quarks and leptons in the
same footing, like composite models \cite{comp,af}, technicolor \cite{tec},
and grand unified theories \cite{gut}, predict the existence of new particles,
called leptoquarks, that mediate quark-lepton transitions. Since leptoquarks
couple to a lepton and a quark, they are color triplets under $SU(3)_C$, carry
simultaneously lepton and baryon number, have fractional electric charge, and
can be of scalar or vector nature.

From the experimental point of view, leptoquarks possess the striking
signature of a peak in the invariant mass of a charged lepton with a jet,
which make their search rather simple without the need of intricate analyses
of several final state topologies.  So far, all leptoquark searches led to
negative results. At the hadron colliders, leptoquarks can be pair produced by
gluon--gluon and quark--quark fusions, as well as singly produced in
association with a lepton in gluon--quark reactions. At the Tevatron, the CDF
and D\O\ collaborations studied the pair production of leptoquarks which decay
into electron-jet pairs \cite{PP}. The combined CDF and D\O\ limit on the
leptoquark mass is $M_{\text{lq}} > 242$ GeV \cite{pp:comb} for scalar
leptoquarks decaying exclusively into $e^\pm$--jet pairs.  At HERA, first
generation leptoquarks are produced in the $s$--channel through their Yukawa
couplings, and the HERA experiments \cite{HERA} placed limits on their masses
and couplings, establishing that $M_{lq} \gtrsim 215-275$ GeV depending on the
leptoquark type.

In this work we studied the single production of first generation leptoquarks
($S$) in association with a charged lepton at the Tevatron \cite{th:pp}, {\em
i.e.}
\begin{equation}
	p \bar{p} \to S~ e^\pm \to e^+ e^- \hbox{ jets} \;\; .
\end{equation}
We performed a careful analyses of all possible QCD and electroweak
backgrounds for the topology exhibiting jets plus a $e^+e^-$ pair using the
event generator PYTHIA \cite{pyt}. The signal was also generated using this
package.  We devised a series of cuts not only to reduce the background, but
also to enhance the signal. Since the available phase space for single
production is larger than the one for double production, we show that the
single leptoquark search can extend considerably the Tevatron bounds on these
particles. Our results indicate that the combined results of the Tevatron
experiments can exclude the existence of leptoquarks with masses up to
260--285 (370--425) GeV at the RUN I (RUN II), depending on their type, for
Yukawa couplings of the electromagnetic strength.

It is interesting to notice that pair production of scalar leptoquarks in a
hadronic collider is essentially model independent since the leptoquark--gluon
interaction is fixed by the $SU(3)_C$ gauge invariance. On the other hand,
single production is model dependent because it takes place via the unknown
leptoquark Yukawa interactions.  Notwithstanding, these two signals for scalar
leptoquarks are complementary because they allow us not only to reveal their
existence but also to determine their properties such as mass and Yukawa
couplings to quarks and leptons. In this work, we also studied the region in
the parameter space where the single leptoquark production can be isolated
from the pair production.

The outline of this paper is as follows. In Sec.\ \ref{l:eff} we introduce the
$SU(3)_C \otimes SU(2)_L \otimes U(1)_Y$ invariant effective Lagrangians that
we analyzed. We also discuss in this section the main features of the
leptoquark signal and respective backgrounds.  We present our results in Sec.\
\ref{resu} while our conclusions are drawn in Sec.\ \ref{conc}.

\section{Leptoquark signals and backgrounds}
\label{l:eff}

We assumed that scalar--leptoquark interactions are $SU(3)_C \otimes SU(2)_L
\otimes U(1)_Y$ gauge invariant above the electroweak symmetry breaking scale
$v$.  Moreover, leptoquarks must interact with a single generation of quarks
and leptons with chiral couplings in order to avoid the low energy constraints
\cite{shanker,fcnc}. The most general effective Lagrangian satisfying these
requirements and baryon number (B), lepton number (L), electric charge, and
color conservations is \cite{buch}
\begin{eqnarray}
{\cal L}_{{eff}}~  &=& {\cal L}_{F=2} ~+~ {\cal L}_{F=0}  \;\; , 
\label{e:int}
\\
{\cal L}_{F=2}~  &=& g_{{1L}}~ \bar{q}^c_L~ i \tau_2~ 
\ell_L ~S_{1L}+ 
g_{{1R}}~ \bar{u}^c_R~ e_R ~ S_{1R} 
+ \tilde{g}_{{1R}}~ \bar{d}^c_R ~ e_R ~ \tilde{S}_1
\nonumber \\
&& +~ g_{3L}~ \bar{q}^c_L~ i \tau_2~\vec{\tau}~ \ell_L \cdot \vec{S}_3 
\;\; ,
\label{lag:fer}
\\
{\cal L}_{F=0}~  &=& h_{{2L}}~ R_{2L}^T~ \bar{u}_R~ i \tau_2 ~
 \ell_L 
+ h_{{2R}}~ \bar{q}_L  ~ e_R ~  R_{2R} 
+ \tilde{h}_{{2L}}~ \tilde{R}^T_2~ \bar{d}_R~ i \tau_2~ \ell_L
\;\; ,
\label{eff} 
\end{eqnarray}
where $F=3B+L$, $q$ ($\ell$) stands for the left-handed quark (lepton)
doublet, and we omitted the flavor indices of the leptoquark couplings to
fermions.  The leptoquarks $S_{1R(L)}$ and $\tilde{S}_1$ are singlets under
$SU(2)_L$, while $R_{2R(L)}$ and $\tilde{R}_2$ are doublets, and $S_3$ is a
triplet.

From the above interactions we can see that first generation leptoquarks can
decay into pairs $e^\pm q$ and $\nu_e q^\prime$, thus, giving rise to a
$e^\pm$ plus a jet, or a jet plus missing energy. However, the branching ratio
of leptoquarks into these final states depends on the existence of further
decays, {\em e.g.}  into new particles.  In this work we considered only the
$e^\pm q$ decay mode and that the branching ratio into this channel ($\beta$)
is a free parameter.  As we can see from Eqs.\ (\ref{lag:fer}) and
(\ref{eff}), only the leptoquarks $R^2_{2L}$, $\tilde{R}^2_2$ and $S_3^-$
decay exclusively into a jet and a neutrino, and consequently can not give
rise to the topology that we are interested in.

The event generator PYTHIA assumes that the leptoquark interaction with quarks
and leptons is described by
\begin{equation}
\bar{e}~ (a+b\gamma_5)~ q  \;\;,
\end{equation}
and the leptoquark cross sections are expressed in terms of the parameter
$\kappa$ defined as
\begin{equation}
\kappa~ \alpha_{\text{em}} ~\equiv~ \frac{a^2 + b^2}{4\pi}
\label{eq:kap}
\end{equation}
with $\alpha_{\text{em}}$ being the fine structure constant.  We present our
results in terms of the leptoquark mass $M_{lq}$ and $\kappa$, being trivial
to translate $\kappa$ into the coupling constants appearing in Eqs.\
(\ref{lag:fer}) and (\ref{eff}); see Table \ref{t:cor}.  The subprocess cross
section for the associated production of a leptoquark and a charged lepton
\begin{equation}
q + g \rightarrow S + \ell \;\; , 
\end{equation}
depends linearly on the parameter $\kappa$ defined in Eq.\ (\ref{eq:kap}). For
the range of leptoquark masses accessible at the Tevatron, leptoquarks are
rather narrow resonances with their width given by
\begin{equation}
	\Gamma( S \to \ell q) = \frac{\kappa \alpha_{\text{em}}}{2}
	M_{\text{lq}} \;\; .
\end{equation}

At the parton level, the single production of leptoquarks leads to a final
state exhibiting a pair $e^+e^-$ and $q$ ($\bar{q}$).  After the parton shower
and hadronization the final state usually contains more than one jet. An
interesting feature of the final state topology $e^+e^-$ and jets is that the
double production of leptoquarks also contribute to it. Consequently, the
topology $e^+e^-$--jets has a cross section larger than the pair or single
leptoquark productions alone, increasing the reach of the Tevatron.  In
principle we can separate the single from the double production, for instance,
requiring the presence of a single jet in the event. However, in the absence
of any leptoquark signal, it is interesting not to impose this cut since in
this case the signal cross section gets enhanced, leading to more stringent
bounds.

We exhibit in Table \ref{t:sin1} the total cross section for the single
production of leptoquarks that couple only to $e^\pm u$ or $e^\pm d$ pairs,
assuming $\kappa=1$ and $\beta=1$ and requiring one electron with $p_T > 50$
GeV, another one with $p_T > 20$ GeV, and $|\eta|<4.2$ for both $e^\pm$.
Notice that the cross sections for the single production of $e^+q$ and $e^-q$
leptoquarks, that is $|F|=0$ or $2$, are equal at the Tevatron.  Furthermore,
the cross section for the single production of a leptoquark coupling only to
$u$ quarks is approximately twice the one for leptoquarks coupling only to $d$
quarks, in agreement with a naive valence--quark--counting rule.  We display
in Table \ref{t:dou} the production cross section of leptoquark pairs for the
same choice of the parameters and cuts used in Table \ref{t:sin1}. The small
difference between the cross sections for the production of $e^\pm u$ and
$e^\pm d$ leptoquarks is due to the exchange of a $e^\pm$ in the $t$--channel
of the reaction $q \bar{q} \to S
\bar{S}$.

In our analyses we kept track of the $e^\pm$ (jet) carrying the largest
transverse momentum, that we denoted by $e_1$ ($j_1$), and the $e^\pm$ (jet)
with the second largest $p_T$, that we called $e_2$ ($j_2$).  The
reconstruction of the jets in the final state was done using the subroutine
LUCELL of PYTHIA, requiring the transverse energy of the jet to be larger than
7 GeV inside a cone $\Delta R = \sqrt{ \Delta \eta^2 + \Delta \phi^2} =0.7$.

\begin{figure}
\begin{center}
\mbox{\qquad\epsfig{file=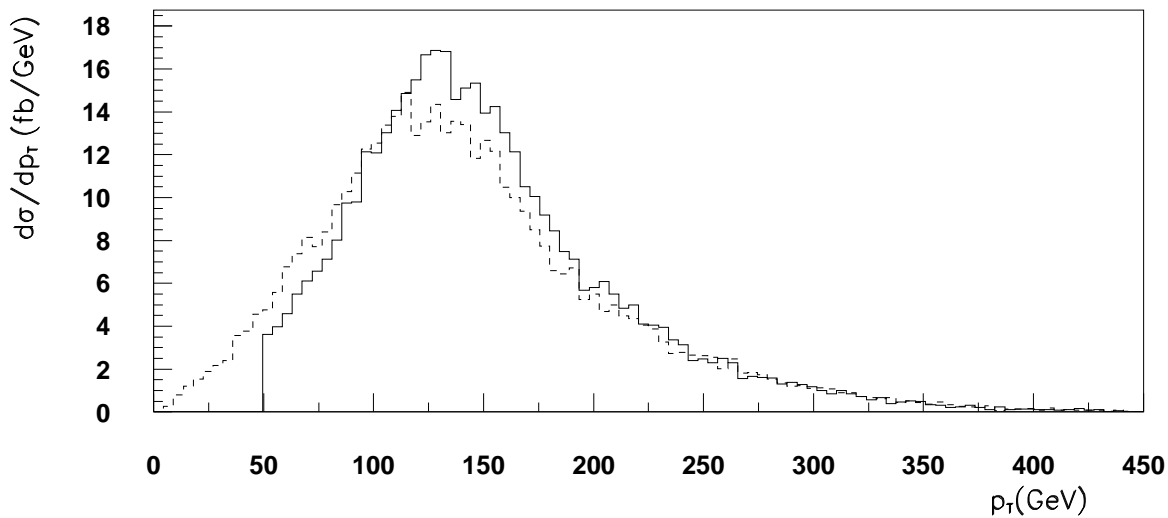,width=\linewidth}}
\end{center}
\caption{$p_T$ spectrum of the largest transverse momentum $e^\pm$ (solid line)
and jet (dashed line). We added the single and double productions of $u e^+$
leptoquarks with mass $M_{\protect\text{lq}}=300$ GeV for
$\protect\sqrt{s}=2.0$ TeV, $\kappa=1$, and $\beta=1$. }
\label{sig:pt}
\end{figure}

The transverse momentum distributions of the $e_{1}$ and $j_{1}$ originating
from leptoquarks are shown in Fig.\ \ref{sig:pt}, where we required that
$p_T^{e_1} > 50$ GeV, $p_T^{e_2} > 20$ GeV, and $|\eta|<4.2$ for both
$e^\pm$. In this figure, we added the contributions from single and pair
production of $u e^-$ leptoquarks of mass $M_{\text{lq}}=300$ GeV for
$\sqrt{s}=2.0$ TeV.  We can see from this figure that the $e_1$ and $j_1$
spectra are peaked approximately at $M_{\text{lq}}/2$ and exhibit a large
fraction of hard leptons, and consequently the $p_T$ cut on $e_1$ does not
reduce significantly the signal.

Within the scope of the SM, there are many sources of backgrounds leading to
jets accompanied by a pair $e^+e^-$. We divided them into three classes
\cite{nos}:

\begin{itemize}

\item {\em QCD processes:} The reactions included in the QCD class are
initiated by hard scatterings proceeding exclusively through the strong
interaction. In this class of processes, the main source of hard $e^\pm$ is
the semileptonic decay of hadrons possessing quarks $c$ or $b$, which are
produced in the hard scattering or in the parton shower through the splitting
$g \rightarrow c \bar{c}$ ($b \bar{b}$).  Important features of the events in
this class are that close to the hard $e^\pm$ there is a substantial amount of
hadronic activity and that the $e^\pm$ transverse momentum spectrum is peaked
at small values.

\item{\em Electroweak processes:} This class contains the Drell-Yan production
of quark/lepton pairs and the single and pair productions of electroweak gauge
bosons.  It is interesting to notice that the main backgrounds by far in this
class are $q_i g~ (\bar{q_i}) \rightarrow Z q_i~ (g) $. This suggests that we
should veto events where the invariant mass of the $e^+e^-$ pair is around the
$Z$ mass; however, even after such a cut, these backgrounds remain important
due to the production of off-shell $Z$'s.

\item{\em Top production:} The production of top quark pairs takes place
through quark--quark and gluon--gluon fusions. In general, the $e^\pm$
produced in the leptonic top decay into $be\nu_e$ are rather isolated and
energetic.  Fortunately, the top production cross section at the Tevatron is
rather small.

\end{itemize}

As an illustration, we present in Table \ref{t:bckg} the total cross section
of the above background classes requiring the events to exhibit a $e^\pm$ with
$p_T>50$ GeV and a second $e^\mp$ having $p_T>20$ GeV with the invariant mass
of this pair differing from the $Z$ mass by more than 5 GeV.  As we can see
from this table, the introduction of this $p_T$ cut already reduces the QCD
backgrounds to a level below the electroweak processes without on--mass-shell
production of $Z$'s.  As we naively expect, the increase in the
center--of--mass energy has a great impact in the top production cross
section.

\section{Results}
\label{resu}

Taking into account the features of the signal and backgrounds, we imposed the
following set of cuts:

\begin{itemize}

\item [(C1)] We required the events to exhibit a pair $e^+e^-$ and one or more
jets.

\item [(C2)] We introduced typical acceptance cuts -- that is, the $e^\pm$ are
required to be in the rapidity region $|\eta_e| < 2.5$ and the jet(s) in the
region $|\eta_j|<4.2$.

\item [(C3)] One of the $e^\pm$ should have $p_T>50$ GeV and the other
$p_T>20$ GeV.

\item [(C4)] The $e^\pm$ should be isolated from hadronic activity. We
required that the transverse energy deposited in a cone of size $\Delta R =0.5$
around the $e^\pm$ direction to be smaller than 10 GeV.

\item [(C5)] We rejected events where the invariant mass of the pair $e^+e^-$
($M_{e_1 e_2}$) is close to the $Z$ mass, {\em i.e.}  we demanded that
$|M_{e_1 e_2} - M_Z| < 30$ GeV. This cut reduces the backgrounds coming from
$Z$ decays into a pair $e^+ e^-$.

\item [(C6)] We required that {\em all} the invariant masses $M_{e_i j_k}$
($i$, $k=1$, $2$) are larger than 10 GeV.

\item [(C7)] We accepted only the events which exhibit a pair $e^\pm$--jet
with an invariant mass $M_{ej}$ in the range $|M_{ej}-M_{\text{lq}}|< 30$ GeV.
An excess of events signals the production of a leptoquark.

\end{itemize}

\begin{figure}
\begin{center}
\mbox{\qquad\epsfig{file=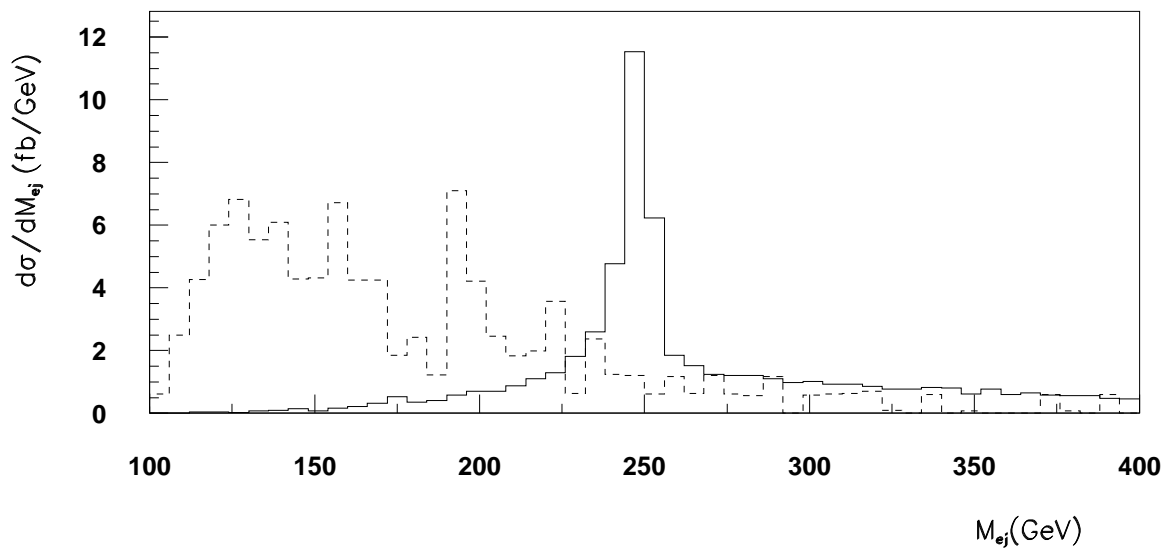,width=\linewidth}}
\end{center}
\caption{$M_{ej}$ spectrum due to the background (dashed line) 
and a leptoquark of mass 250
GeV with $\kappa=1$ and $\beta=1$ (solid line) after the cuts
(C1)--(C6)  are applied for $\protect\sqrt{s}=1.8$~ TeV.}
\label{fig:mej}
\end{figure}

In Fig.\ \ref{fig:mej} we present the $M_{ej}$ spectrum after the cuts
(C1)--(C6) originating from the background and the production of a $e^+ u$
leptoquark of mass 250 GeV with $\kappa= \beta = 1$. The largest invariant
mass of the four possible combinations is plotted both for background (dashed
line) and signal (solid line). The signal peak is clearly seen out of the
background.


\subsection{Pair production}

At this point it is interesting to obtain the attainable bounds on leptoquarks
springing from the search of leptoquark pairs. In this case we required, in
addition to cuts (C1)--(C7), that the events present two $e^\pm$--jet pairs
with invariant masses satisfying $|M_{ej}-M_{\text{lq}}|< 30$ GeV. Our results
show that CDF and D\O\ should be able to constrain the leptoquark masses to be
heavier than 225 (350) GeV at the RUN I (RUN II) for $\beta=1$ and $\kappa=0$,
assuming that only the background is observed. When the data of both
experiment are combined, the limit changes to 250 (375) GeV. It is interesting
to notice that our results for the RUN I are compatible with the ones obtained
by the Tevatron collaborations \cite{nota}. Moreover, taking into account the
single production of leptoquarks changes these constraints only by a few GeV
for $\kappa=1$.


\subsection{Single production}

\begin{figure}
\begin{center}
\parbox[c]{3.in}{
\mbox{\qquad\epsfig{file=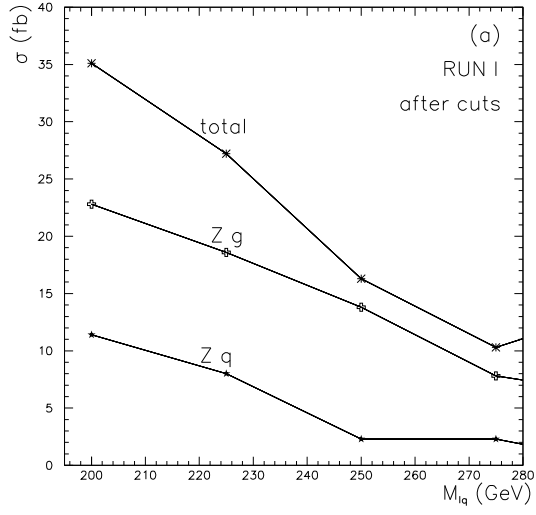,width=\linewidth}}
}
\hfill
\parbox[c]{3.in}{
\mbox{\qquad\epsfig{file=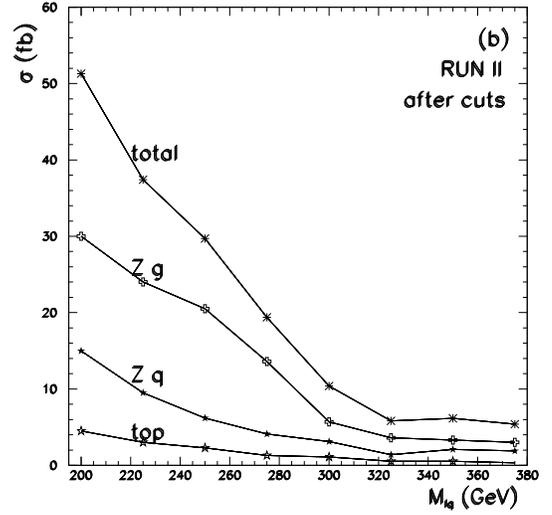,width=\linewidth}}
}
\end{center}
\caption{Total cross sections of the main backgrounds after cuts for (a)
$\protect\sqrt{s}=$ 1.8 and (b) 2.0 TeV. The line labeled $Z g$ ($Z q$) stands
for the reaction $q \bar{q} \to Z g$ ($q g \to Z q$) while the line marked top
represents the cross section for the production of top pairs.}
\label{fig:xback}
\end{figure}

\begin{figure}
\begin{center}
\parbox[c]{3.in}{
\mbox{\qquad\epsfig{file=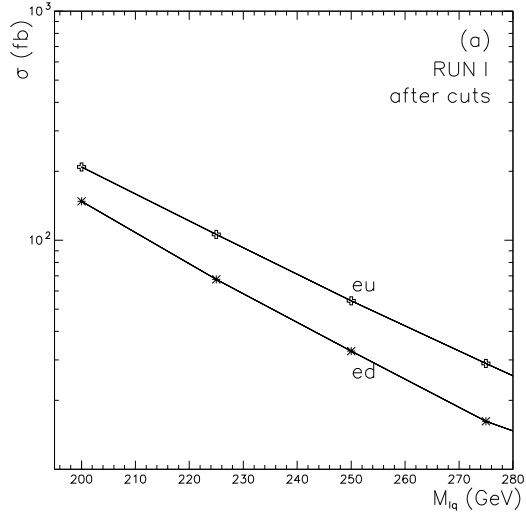,width=\linewidth}}
}
\hfill
\parbox[c]{3.in}{
\mbox{\qquad\epsfig{file=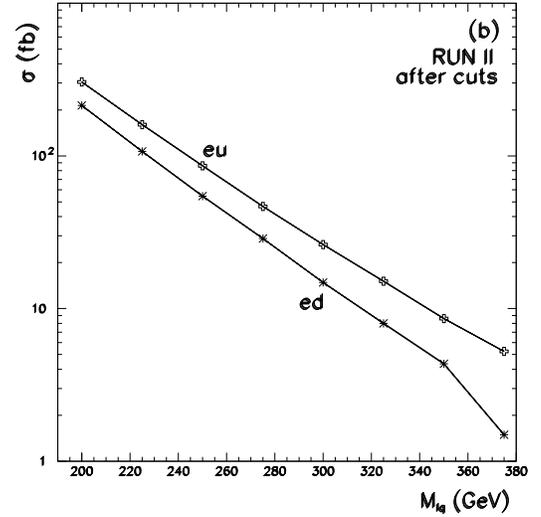,width=\linewidth}}
}
\end{center}
\caption{Total cross sections for the production of $e^+ u$ and $e^+ d$
leptoquarks after cuts for (a) $\protect\sqrt{s}=$ 1.8 and (b) 2.0 TeV.  We
assumed $\kappa=1$ and $\beta=1$.}
\label{fig:xsign}
\end{figure}

We display in Fig.\ \ref{fig:xback} the total background cross section and its
main contributions as a function of $M_{lq}$ after applying the cuts
(C1)--(C7) for center--of--mass energies of 1.8 and 2.0 TeV. We can see from
this figure that the number of expected background events per experiment at
the RUN I (II) is 4 (102) for $M_{\text{lq}}=200$ GeV dropping to 0 (8) for
$M_{\text{lq}}=400$ GeV.  For the sake of comparison, we display in Fig.\
\ref{fig:xsign} the total cross section for the production of $e^+ u$ and $e^-
d$ leptoquarks assuming $\kappa=1$ and $\beta=1$ for the same cuts and
center--of--mass energies.

\begin{figure}
\begin{center}
\parbox[c]{3.in}{
\mbox{\qquad\epsfig{file=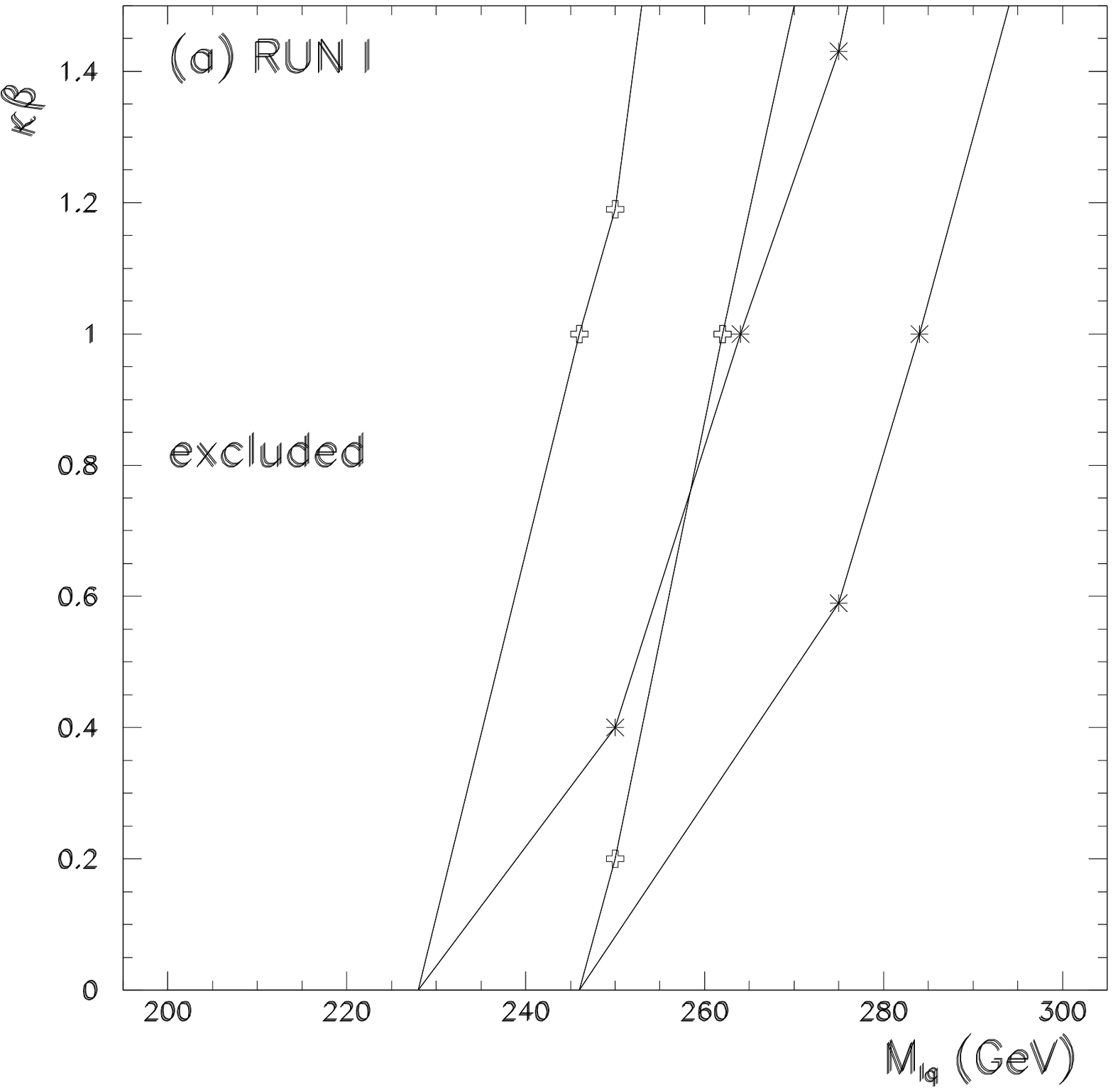,width=\linewidth}}
}
\hfill
\parbox[c]{3.in}{
\mbox{\qquad\epsfig{file=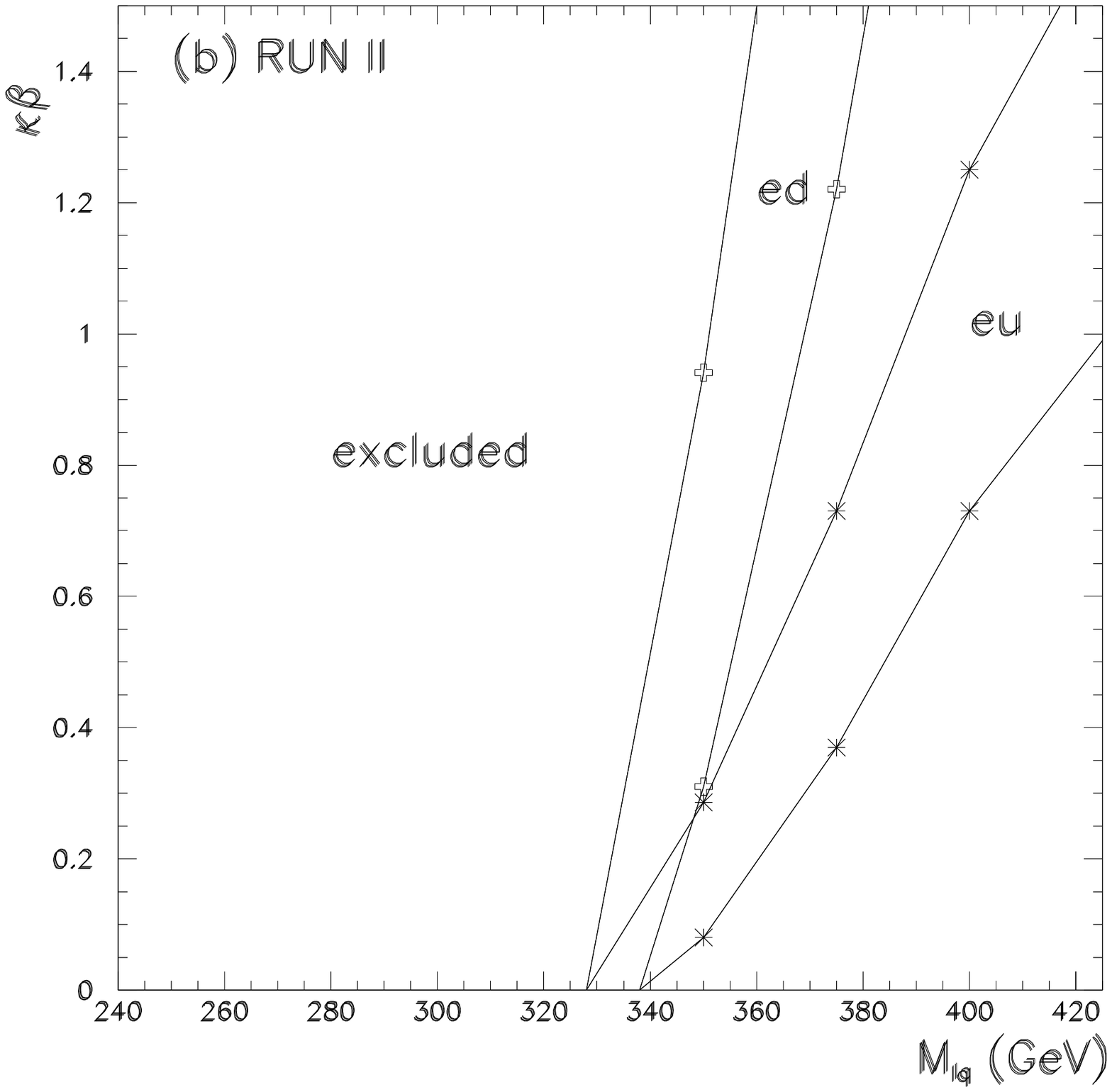,width=\linewidth}}
}
\end{center}
\caption{95\% CL excluded region in the $\kappa\beta \otimes 
M_{\protect\text{lq}}$ for (a) $\protect\sqrt{s}=$ 1.8 and (b) 2.0 TeV. The
curves with crosses (stars) correspond to the bounds on $e^\pm d$ ($e^\pm u$)
leptoquarks, with the upper (lower) one originating from the results of a
single (combined) experiment(s).  }
\label{fig:res}
\end{figure}

We estimated the capability of the Tevatron to exclude regions of the plane
$\kappa\beta \otimes M_{\text{lq}}$ assuming that only the background events
were observed. We present in Fig.\ \ref{fig:res}a the projected 95\% CL
exclusion region for $e^+ u$ and $e^+ d$ leptoquarks at the RUN I with an
integrated luminosity of 110 pb$^{-1}$ per experiment. From our results we can
learn that the search for single $e^\pm u$ ($e^\pm d$) leptoquarks in each
experiment can exclude leptoquark masses up to 265 (245) GeV for $\kappa
\beta=1$. Combining the results of CDF and D\O\ expands this range of excluded
masses to 285 (260) GeV respectively.  The corresponding results for the RUN
II with an integrated luminosity of 2 fb$^{-1}$ per experiment are presented
in Fig.\ \ref{fig:res}b. Here we can see that the combined CDF and D\O\ data
will allow us to rule out $e^\pm u$ ($e^\pm d$) leptoquarks with masses up to
425 (370) GeV, assuming that $\beta=\kappa=1$.

It is important to stress that events exhibiting a pair of leptoquarks also
contribute to our single leptoquark search. This is the reason why lighter
leptoquarks can be observed even for arbitrarily small $\kappa$; see Figs.\
\ref{fig:res}. However, the maximum mass that can be excluded for $\kappa = 
0$ is smaller than the limit coming from the search for leptoquark pairs since
the requirement of an additional $e^\pm$--jet pair with invariant mass close
to $M_{\text{lq}}$ helps to further reduce the backgrounds. For instance the
single leptoquark search can rule out leptoquarks with masses up to 330 GeV
for $\kappa=0$ at RUN II while the search for leptoquark pairs leads to a
lower bound of 375 GeV.

\begin{figure}
\begin{center}
\mbox{\qquad\epsfig{file=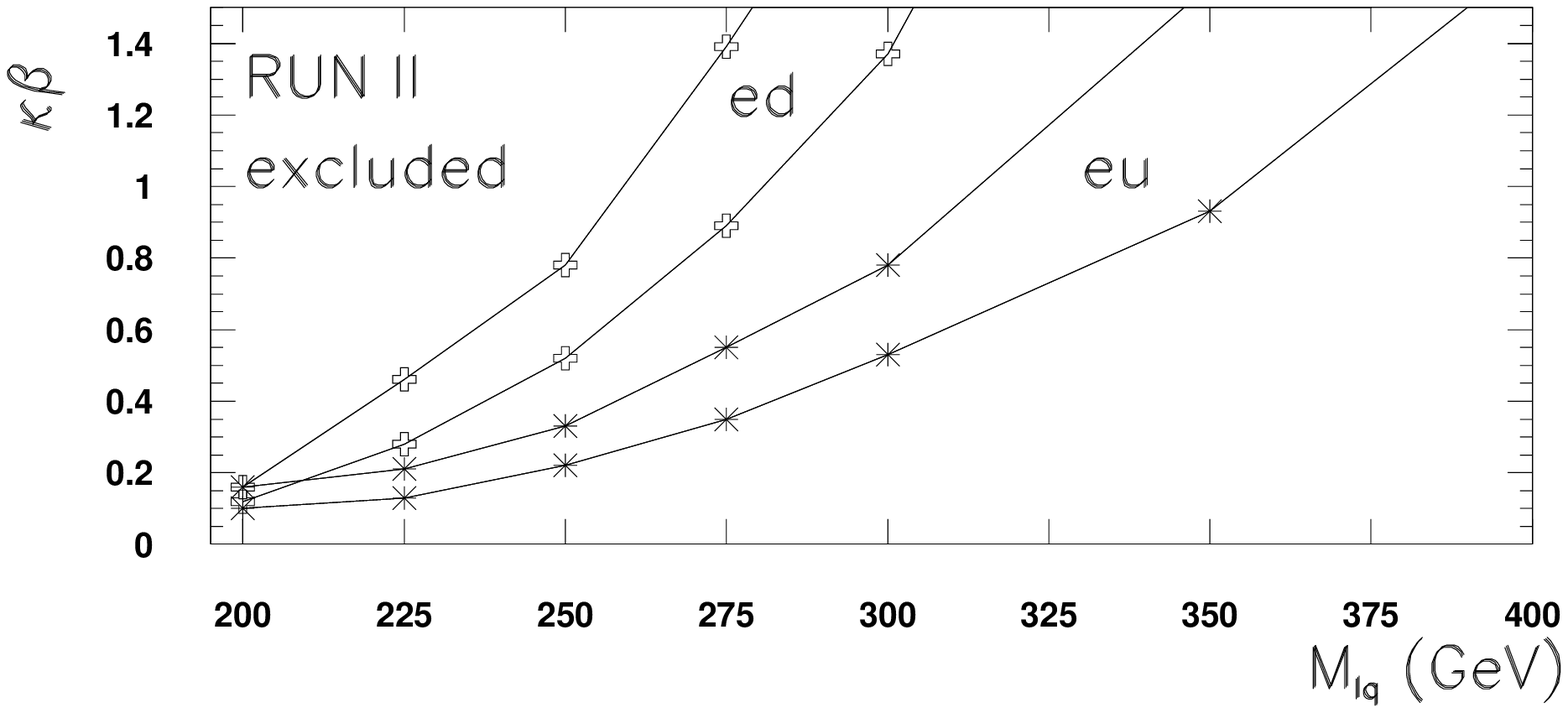,width=\linewidth}}
\end{center}
\caption{95\% CL excluded region in the $\kappa\beta \otimes 
M_{\protect\text{lq}}$ for $\protect\sqrt{s}=$ 2.0 TeV when we impose cuts
(C1)--(C7) and also demand that the events exhibit just one jet. The curves
with crosses (stars) correspond to the bounds on $e^\pm d$ ($e^\pm u$)
leptoquarks, with the upper (lower) one originating from the results of a
single (combined) experiment(s).  }
\label{fig:sing}
\end{figure}

In principle we can separate the double production of leptoquarks from the
single one. An efficient way to extract the single leptoquark events is to
require that just one jet is observed. At the RUN I this search leads to an
observable effect only for rather large values of $\kappa$. On the other hand,
this search can be done at the RUN II, however, the bounds coming from this
analysis are weaker than the ones obtained above; see Fig.\ \ref{fig:sing}.
We can interpret this figure as the region of the $\kappa \beta \otimes
M_{\text{lq}}$ where we can isolat the single leptoquark production and
study this process in detail.

\section{Conclusions}
\label{conc}

The analyses of the single production of leptoquarks at the Tevatron RUN I
allow us to extend the range of excluded masses beyond the present limits
stemming from the search of leptoquark pairs. We showed that in the absence of
any excess of events CDF and D\O\ individually should be able to probe $e^\pm
u$ ($e^\pm d$) leptoquark masses up to 265 (245) GeV for Yukawa couplings of
the electromagnetic strength and $\beta=1$. In the case $\beta=0.5$ these
limits reduce to 250 (235) GeV. Moreover, combining the results from both
experiments can further increase the Tevatron reach for leptoquarks.  Assuming
that leptoquarks decay exclusively into the known quarks and leptons and
$\kappa=1$, the combined Tevatron results can exclude $S_{1L}$ and $S_3^0$
leptoquarks with masses up to 270 GeV, $S_{1R}$, $R^1_{2L}$, and $R^1_{2R}$
leptoquarks with masses 285 GeV, and $\tilde{S}_{1R}$, $S^+_3$, $R^2_{2R}$,
and $\tilde{R}^1_2$ with masses up to 260 GeV. This results represent an
improvement over the present bounds obtained at the Tevatron \cite{PP},
however, the bounds are similar to the ones obtained by the HERA
collaborations \cite{HERA}.

At the RUN II, the search for the single production of leptoquarks will be
able to rule out leptoquarks with masses even larger. For instance, the CDF
and D\O\ combined results can probe $e^\pm u$ ($e^\pm d$) leptoquark masses up
to 425 (370) GeV for $\kappa\beta=1$. In the case $\kappa\beta=0.5$, these
bounds reduce to 385 (350) GeV. However, even these improved limits will not
reach the level of the indirect bounds ensuing from low energy physics
\cite{shanker,fcnc,low}. Direct limits more stringent than the indirect ones
will only be available at the LHC \cite{nos,fut:pp} or future $e^+e^-$
colliders \cite{fut:ee}.


\section*{Acknowledgments}

This work was supported by Conselho Nacional de Desenvolvimento
Cient\'{\i}fico e Tecnol\'ogico (CNPq), by Funda\c{c}\~ao de Amparo \`a
Pesquisa do Estado de S\~ao Paulo (FAPESP), and by Programa de Apoio a
N\'ucleos de Excel\^encia (PRONEX).



\begin{table}[htbp]
\begin{center}
\begin{tabular}{|c|c|c|c|}
leptoquark & decay & branching ratio & $4\pi\alpha_{\text{em}}\kappa$
\\
\hline
$S_{1L} $         & $e^+ \bar{u}$ &  50\%  & $\frac{g_{1L}^2}{2}$ \\
$S_{1R} $         & $e^+ \bar{u}$ & 100\%  & $\frac{g_{1R}^2}{2}$ \\
$\tilde{S}_{1R}$  & $e^+ \bar{d}$ & 100\%  & $\frac{\tilde{g}_{1R}^2}{2}$ \\
$S^+_3$           & $e^+ \bar{d}$ & 100\%  & $g_3^2$ \\
$S^0_3$           & $e^+ \bar{u}$ &  50\%  & $\frac{g_3^2}{2}$   \\
$R_{2L}^1$        & $e^- \bar{u}$ & 100\%  & $\frac{h_{2L}^2}{2}$ \\
$R_{2R}^1$        & $e^- \bar{u}$ & 100\%  & $\frac{h_{2R}^2}{2}$ \\
$R_{2R}^2$        & $e^- \bar{d}$ & 100\%  & $\frac{h_{2R}^2}{2}$ \\
$\tilde{R}_2^1$   & $e^- \bar{d}$ & 100\%  & $\frac{\tilde{h}_{2L}^2}{2}$ \\
\end{tabular}
\vskip 0.75cm
\caption{Leptoquarks that can be observed through their decays into a $e^\pm$
and a jet and the correspondent branching ratios into this channel assuming
that there are no new particles. We also show the relation between the
leptoquark Yukawa coupling and the parameter $\kappa$ used in PYTHIA.}
\label{t:cor}
\end{center}
\end{table}

\begin{table}[htbp]
\begin{center}
\begin{tabular} {|c|c|c|c|c|}
$\ell q$ coupling  &  $M_{\text lq}=200$ GeV & 250 GeV & 300 GeV & 350 GeV
\\
\hline
$e^\pm u$ & 187./259. & 59./86 & 20./30 & --/12. 
\\
$e^\pm d$ &  77./112. & 22./34 &  7./11. & --/4. 
\\
\end{tabular}
\vskip 0.75cm
\caption{Total cross section in fb for the single production of a leptoquark
that couples only to $\ell q$ pairs for several leptoquark masses and
center--of--mass energies of 1.8/2.0 TeV. We required that one $e^\pm$ has
$p_T > 50$ GeV, the other one $p_T>20$ GeV, and $|\eta|<4.2$ for both $e^\pm$
and assumed $\kappa\beta=1$. We indicate by -- when the cross section is
negligible }
\label{t:sin1}
\end{center}
\end{table}

\begin{table}[htbp]
\begin{center}
\begin{tabular} {|c|c|c|c|c|}
$\ell q$ coupling  &  $M_{\text lq}=200$ GeV & 250 GeV & 300 GeV & 350 GeV
\\
\hline
$e^\pm u$ & 153./237. & 30.9/53.1 & 6.6/13.0 & --/3.23 
\\
$e^\pm d$ & 153./225. & 29./50.1 & 6.2/12.0 & --/3.00
\\
\end{tabular}
\vskip 0.75cm
\caption{Total cross section in fb for the pair production leptoquarks that
couples only to $\ell q$ pairs for several leptoquark masses and
center--of--mass energies of 1.8/2.0 TeV. We imposed the same cuts as in Table
\protect\ref{t:sin1}. }
\label{t:dou}
\end{center}
\end{table}

\begin{table}[htbp]
\begin{center}
\begin{tabular} {|c|c|c|}
Class        &  $\sigma_{\protect\text{total}}$(1.8 TeV) (fb)
&  $\sigma_{\protect\text{total}}$(2.0 TeV) (fb)
\\
\hline
QCD          &   67.   & 129.
\\
electroweak  &  453.   & 562.
\\
top          &  3.9    &  52.
\\
\end{tabular}
\vskip 0.75cm
\caption{Total cross section in fb of the different background classes for
center--of--mass energies of 1.8 and 2.0 TeV. We required one $e^\pm$ with
$p_T > 50$ GeV and the other $e^\mp$ with $p_T>20$ GeV. We also demanded that
the invariant mass of $e^+ e^-$ pair differs from the $Z$ mass by more than 5
GeV.  }
\label{t:bckg}
\end{center}
\end{table}



\begin{thebibliography}{99}
\bibitem{nota} CDF and D\O\ collaborations use the NLO pair production cross
sections in their analyses, while we use the tree level expression. The
difference in the cross sections compensates the fact that we did not consider
detection efficiencies and detector resolution.
\end{thebibliography}
\end{document}